\documentstyle[12pt,epsfig]{article}
\begin{document}

\setlength{\textwidth}{15truecm}
\setlength{\textheight}{23cm}
\baselineskip=24pt


\bigskip

\centerline {\bf{Parameter Dependence of Transonic Accretion onto a Black Hole}}
\centerline {\bf{in the Presence of Bremsstrahlung and Synchrotron Cooling}}
\medskip
\medskip
\medskip
\centerline {Sivakumar G. Manickam}
\centerline {Harish-Chandra Research Institute,} 
\centerline {Chhatnag Road, Jhunsi, Allahabad 211019, India}
\centerline {email: \it{sivman@mri.ernet.in}}

\ \\
\bigskip

\noindent{\bf ABSTRACT}

\noindent The topologies are obtained for accretion flows around black 
holes as the parameters, accretion rate and specific angular momentum are varied.
Paczy{\'n}ski-Wiita pseudo-Newtonian potential is used to mimic the gravitational field
of a Schwarzschild black hole.
Both stellar mass($14M_{\odot}$) and supermassive($10^8M_{\odot}$) black holes are
considered.
The effects of bremsstrahlung and synchrotron cooling processes on 
the flow dynamics are studied.
Rankine-Hugoniot conditions are checked for possible 
shock formation of solution branches of the topology.
Stability of the branches and the boundary conditions of the flow are likely to decide the 
uniqueness of the solution. 

Running Title: Black Hole Accretion with Bremsstrahlung and Synchrotron cooling 

\bigskip


\newpage 
\noindent  {\bf 1. INTRODUCTION} 

\noindent The presence of accretion disc was
theoretically predicted even before they were observed, as with the case of neutron stars and black holes.
Disc structures are expected in various astrophysical systems
when it has angular momentum. Most of the astrophysical
systems have some angular momentum associated with them. Solar system,
galactic disc and accretion disc are some examples. 
The flow is assumed to be in form of a disc, which basically means that we have an axisymmetric
system. As the matter spirals in, its specific angular momentum is dissipated
and energy is lost in the form of electromagnetic radiation. Hence an accretion disc
becomes a powerful source of luminosity.

 The study of accretion processes around a black hole using the equations of the general theory of
relativity is a mathematically
tedious task. But since the gravitational field is quite strong, a Newtonian description
for the potential is not satisfactory especially close to the compact object. So as a compromise
between the two, one can devise pseudo-Newtonian potentials which can produce the salient features of a
black hole geometry. Also upgrading a numerical code which is written using Newtonian potential,
can be trivially
done by replacing the potential with an appropriate pseudo-potential.
Paczy{\'n}ski and Wiita (1980) introduced a pseudo-potential for
Schwarzschild black hole which produces the positions of marginally stable and marginally bound
orbits correctly and the efficiency approximately same as that of the effective potential
of a Schwarzschild black hole.
However it should be realized that while the gravitational force can be obtained as derivative of the
pseudo-potential, one cannot relate the derivative of the effective potential to the gravitational force.

More refined pseudo-potentials for Schwarzschild black hole geometry are discussed in the literature
(Nowak and Wagoner 1991; Artemova, Bjornsson and Novikov 1996).
A Pseudo-Kerr potential has been suggested by Chakrabarti and Khanna (1992) which is valid only for
small Kerr parameter. There are some other works in the literature which deal extensively with the
usage of pseudo-potentials and discuss their relative merits (e.g. Das and Sarkar 2001; Das 2002; Das 2003; Mukhopadhyay 2002;
Mukhopadhyay 2003). In this paper, we mimic the gravitational effects of a Schwarzschild black hole
using Paczy{\'n}ski-Wiita pseudo-potential.

 Simplifying approximations have to be made to keep the problem tractable. One has to make a judicious
choice motivated by application to a physical situation. Zeroth-order approximation is to
assume that the disc has axisymmetry
and it is in steady state. This results in reduction of the problem from (3+1)D to 2D.
The equations can be vertically averaged (for e.g. Shakura and Sunyaev 1973; Chakrabarti 1989; Narayan and Yi 1994),
assuming that the flow to be in hydrostatic equilibrium in the transverse
direction which results in a hybrid 1.5D model.
The assumption of transverse equilibrium is justified
if the sound crossing time in the transverse direction is much less than the infall time,
which in turn requires the disc to be thin.
The results of numerical simulations of Molteni, Lanzafame and Chakrabarti (1994) agree with
vertically averaged solution of Chakrabarti (1989). This implies that vertical averaging is a
valid approximation.

Thin disc of constant transverse height approximation results in a 1D problem which can be used for
testing (2+1)D numerical code (e.g. Chakrabarti and Molteni 1993; Molteni, Sponholz and Chakrabarti 1996).
The thin disc approximation is consistent with the requirement that the flow is highly
supersonic w.r.t. angular velocity $v_\phi$ (e.g. Frank, King and Raine 1985; Longair 1994).
In this work, we model the disc as thin and of constant transverse height.
We also assume that the self-gravity of the accretion disc is negligibly small and hence the gravitational
force experienced by an accreting element is essentially due to that of the central compact object.
The magnetic field does not enter the dynamics of the
flow as it is assumed to be stochastic.

 The global solution of the accretion disc equations has been posed as a challenging 
numerical boundary value problem
by Narayan et al.(1997) where a relaxation method is used.
In this paper,
fourth-order Runge-Kutta method (Press et al. 1992) is used to solve the accretion
disc equations.
As Runge-Kutta method
is fairly easy
to implement, it allows to perform a rigorous and systematic study of the flow topologies, as
the parameters are varied.
Unless one finds a physical context where either one
of these techniques is superior to the other, both are equally acceptable.

\ \\
\noindent {\bf 2. THE BASIC FLOW EQUATIONS}

\noindent The equations which govern the flow are the basic conservation equations of mass, momentum and
energy. The pseudo-Newtonian potential of Paczy{\'n}ski and Wiita (1980) is used, which mimics the
gravitational field around a Schwarzschild black hole to a sufficient accuracy.
To describe the flow, cylindrical coordinate system, $(r, \phi, z)$ is used.
${\bf v}(v_r, v_{\phi}, v_z),$ $\rho$ and $p$ are the velocity, density and pressure of the
accreting matter respectively.

\noindent Continuity equation:

$$
{{{\partial \rho}\over{\partial t}} +{ \nabla .{({\rho}{\bf v})}} = 0 }.
\eqno(1a)
$$

\noindent Euler equation:

$$
{{\partial \bf v} \over {\partial t}} + (\bf v. \nabla ) \bf v + {\nabla p \over \rho} + \nabla g
= 0,
\eqno(1b)
$$

\noindent where, $ g = -{GM \over (r-r_g)} $ is the Paczy{\'n}ski-Wiita potential, $G$ is the gravitational
constant, M is the mass of the black hole, $r_g={2GM \over c^2}$ is the Schwarzschild radius
and $c$ is the velocity of light.

\newpage
\noindent Energy equation:

$$
{\partial \over {\partial t}} (\rho \epsilon) +  \nabla . ( \rho \epsilon \bf v) + \Lambda -
\Gamma = 0,
\eqno(1c)
$$

\noindent where, $ \epsilon = {1 \over 2} {v_r}^2 + U + {p \over \rho} + g + {1 \over 2} {v_ \phi}^2 $
is the specific energy, $U={p \over \rho (\gamma-1)}$ is the thermal energy, $\gamma$ is the
adiabatic index, $\Lambda$ is the cooling term, and $\Gamma$ is the heating term.

\ \\
\noindent {\bf 3. SOLUTION OF TRANSONIC FLOW WITH BREMSSTRAHLUNG AND SYNCHROTRON COOLING}

\noindent We consider steady axisymmetric flow and use the approximation that the disc is thin and of constant
height. Hence, velocity field of
the flow is two-dimensional in the $r \phi$ plane. The dynamical equations of the flow
are the conservation equations of mass, momentum and energy.
The cooling effects due to bremsstrahlung and synchrotron radiation processes are included 
in the energy equation.
In an electron-proton plasma, the expression for bremsstrahlung cooling process
(Lang 1980) is given as,

$$
\Lambda_{brems}=1.43 \times 10^{-27} N_e N_i T^{1/2} Z^2 g_f  \ \ erg
 \ cm^{-3} \ s^{-1},
$$

\noindent where,

$$
N_i Z = {\rho \over m_p + m_e} \approx {\rho \over m_p},
$$

\noindent i.e.,

$$
\Lambda_{brems}=1.43 \times 10^{-27} {\rho ^2 \over m_p ^2} T^{1/2} g_f,
\eqno(2)
$$

\noindent and for synchrotron cooling process (Shapiro and Teukolsky 1983) the expression is,

$$
\Lambda_{sync} = {16 \over 3}{e^2 \over c} ({eB \over {m_e c}})^2 ({kT\over{m_ec^2}})^2 N_e \ \ erg
 \ cm^{-3} \ s^{-1}.
$$

\noindent The equipartition value for the magnetic field($B$), which is obtained by equating the magnetic
energy per
unit mass to the thermal energy per unit mass is given as,

$$
B = \sqrt{ {4\pi \over \gamma(\gamma-1)} a^2\rho }.
\eqno(3)
$$

\noindent Using the ideal gas equation, we get for the temperature $T$,

$$
T = {p \over \rho} \ {\mu m_p \over k} = {a^2 \over \gamma} {\mu m_p \over k}.
\eqno(4)
$$

\noindent Using the above expressions for $ B$ and $T$ we get,

$$
\Lambda_{sync} = {16 \over 3}{e^2 \over c} ({e \over {m_e c}})^2 {4\pi \over {\gamma(\gamma-1)}}
a^2 \rho ({1\over{m_ec^2}})^2 ({a^2 \mu m_p \over \gamma})^2 {\rho \over m_p},
$$

\noindent i.e.,

$$
\Lambda_{sync}
= {16 \over 3}{e^2 \over c}({e \over {m_e c}})^2  {4\pi \over {\gamma(\gamma-1)}} ({1\over{m_ec^2}})^2
({\mu m_p \over \gamma})^2 {1 \over m_p} { a^6 \rho ^2 },
\eqno(5)
$$

\noindent where, $N_e$ and $N_i$ are electron and ion number densities respectively, $Z$ is the atomic number,
$m_p$ is the mass
of proton, $g_f$ is the Gaunt factor, $e$ is the electron charge,
$m_e$ is the electron mass and $\mu=0.5$ for purely hydrogen gas.

\noindent The continuity equation:

$$
{d \over {dr}} (\rho v r) = 0.
\eqno(6a)
$$

\newpage
\noindent The radial momentum equation:

$$
v { dv \over {dr} } + {1 \over \rho} {dp \over {dr}} + \psi ^ \prime = 0,
\eqno(6b)
$$

$$
\psi = -{GM \over (r-r_g)} + {1 \over 2} {v_ \phi}^2 = -{GM \over (r-r_g)} + {1 \over 2} {\lambda^2 \over r^2},
$$

\noindent where, $-{GM \over (r-r_g)} $ is the Paczy{\'n}ski-Wiita potential and $\lambda$ is the specific angular
momentum of the flow.

\noindent The energy equation:

$$
v { dv \over {dr} } + {3 \over 2} \gamma  { d(p/\rho) \over {dr} } + \psi ^ \prime + {{\Lambda_{brems} + \Lambda_{sync}}
\over \rho v} = 0.
\eqno(6c)
$$

The above three equations 6(a-c), are rewritten in a convenient form as,

$$
{dv \over dr} =
{{{\psi ^ \prime - a^2 /r +j{g_f}{2 \over 3} {\rho \over v {m_p}^2} T^{1/2} +s {2 \over 3}
{a^6 \rho \over v}}} \over {a^2 /v -v}},
\eqno(7a)
$$

$$
{1 \over \rho} {d \rho \over dr} + {1 \over v} {dv \over dr} + {1 \over r} = 0,
\eqno(7b)
$$

\noindent and,

$$
v{dv \over dr} + {2a \over \gamma}{da \over dr}+{a^2 \over \rho \gamma}{d\rho \over dr}
+\psi ^ \prime=0,
\eqno(7c)
$$

\noindent where, $^\prime$ denotes the derivative with respect to $r$, $v=v_r$,
$j=1.43 \times 10^{-27}$ in cgs units, $s={16 \over 3}{e^2 \over c}({e \over {m_e c}})^2
{4\pi \over {\gamma(\gamma-1)}} ({1\over{m_ec^2}})^2 ({\mu m_p \over \gamma})^2 {1 \over m_p}$
and the polytropic relation $p=K \rho ^ \gamma$ is
used to obtain the sound speed $a=\sqrt {\partial p \over
\partial \rho}=\sqrt{\gamma p \over \rho}$.

The above set of three equations 7(a-c), are solved 
using fourth-order Runge-Kutta method (Press et al. 1992) as done in Chakrabarti (1990,
see also Chakrabarti 1996a).
The error in the fourth-order Runge-Kutta method is of the order of $h^{5}$, where $h$ is the integration step size.
The solution of Narayan et al.(1997) is not complete as it uses only one critical point. Hence
the possibility of the flow forming a shock is eliminated a priori.

 When the flow velocity becomes equal to the sound speed, the denominator in the expression
for ${dv \over dr}$ (equation 7a) becomes zero. For ${dv \over dr}$ to be finite, numerator should also become
zero. At this point, called the critical point, we use l' Hospital's rule to get,

$$
{dv \over dr} = {-B \pm {\sqrt{(B^2-4AC)}} \over 2A},
\eqno(8)
$$

\noindent where,

$$ A = \gamma+1, $$

$$
B = {\gamma \over a} \psi^\prime - {a \over r} - T_1 {\rho \over v} + T_2 {\gamma \over 2a}
({a \over \gamma} - v) + T_3,
$$

\noindent and,

$$
C = \psi^{\prime\prime} + {a^2 \over r^2}
-T_1 {\rho \over x} + T_2 {\gamma \over 2a}(-\psi^\prime + {a^2 \over \gamma r}).
$$

\noindent The expressions for $T_1$, $T_2$ and $T_3$ are,

$$
T_1 = {2 \over 3} j_1 a^{2\alpha-1} + {2 \over 3} s a^5 \ ; \ j_1 = j g_f {1\over {m_p}^2}({\mu m_p
\over \gamma k})^\alpha,
$$

$$
T_2 = {2 \over 3} j_1 \rho 2\alpha a^{2\alpha-2} + {2 \over 3} s 6 a^4 \rho -2{a \over r},
$$

\noindent and,

$$
T_3 = -{2 \over 3} j_1 \rho a^{2\alpha-2} - {2 \over 3} s a^4 \rho,
$$

\noindent where, $k$ is the Boltzmann constant. Now the transonic solution branches are obtained
by integrating from the critical point. For stellar black hole, we choose the mass to be
$14M_{\odot}$, $\gamma=4/3$ as distinct X-ray variability of GRS 1915+105 is interpreted
as due to instabilities in radiation pressure dominated disc (Greiner, Cuby and McCaughrean 2001).
The mass is chosen as $10^8M_{\odot}$ for super-massive black hole, $\gamma=5/3$ for gas pressure
dominated disc. We consider synchrotron cooling process of variable efficiency. The efficiency
of bremsstrahlung cooling is chosen as unity. For a chosen accretion rate and specific angular
momentum($\lambda$), we scan the r-axis from 1.5$r_g$ to 1000$r_g$ to find if it can be a critical
point. In the case of non-dissipative flow as in Chakrabarti (1989), for a given specific
energy there are unique values for inner and outer critical points. When there is energy
dissipation we have some sort of a degree of freedom in choosing the critical points, each one
corresponding to a different specific energy.

\ \\
\noindent {\bf 3.1 The nature of critical points}

\noindent To find the critical point values, the sonic point conditions have to be used
along with other constraints. The critical point, basically could be X-type (saddle) or O-type (centre)
(Chakrabarti 1989).
The value of the sound speed $a_c$ at the critical point location $r_c$ is found by using the value
of the accretion rate.
The following is the `flow chart' (in FORTRAN 77 syntax) which is used for coding.

 (i) choose the value of the mass of the black hole and polytropic index $gamma$

 (ii) set the efficiency factor of bremsstrahlung cooling ($zefac$) and synchrotron 
\indent cooling ($synfac$)

 (iii) choose the value of accretion rate and specific angular momentum

 (iv) increment $r_c$ value from 1.5$r_g$ to 1000$r_g$ in a loop using $ r_c = r_c + .01*r_c**1.5 $

 (v) for a chosen $r_c$ value, find $a_c$ using the expression,

     $ accrate = 2.*pi*r_c *height* a_c * \rho_c $

     \ \ \indent \indent $ = 2.*pi*r_c *height* a_c * (a_c**2/r_c-siprime)*1.5 / (zeta+sync*a_c**5) $

     $ zeta = zefac * j*g_f/(m_p**2) * (mu*m_p/gamma/k)**alpha $

     $ sync = synfac * 16./3.*e*e/c * ((e/m_e/c)**2) * 4.*pi/gamma/(gamma-1.) * 
\indent ((1/m_e/c/c)**2) * ((mu*m_p/gamma)**2)/m_p $

 (vi) use the bisection routine (Press et al. 1992) to find $a_c$ satisfying the above 
\indent expression for $ accrate $

 The plots of $a_c$ $vs$ $r_c$ obtained using the above routine are shown in Fig. 1 and Fig. 2.
These are representative plots showing the
effect of the variation of synchrotron cooling efficiency factor and
specific angular momentum ($\lambda$)
for a super-massive black hole of mass $10^8M_{\odot}$.
The value of the accretion rate is chosen as unity in Eddington units
and the value for the adiabatic index $\gamma$ is 5/3, for a gas pressure dominated accretion disc.
For all the cases in Fig. 1, the value of $\lambda$ is 1.8 in the units of $2GM/c$.
The lowermost curve corresponds to the bremsstrahlung cooling case without any synchrotron
cooling. The synchrotron cooling factor is increased as 0, $10^{-7}$, $10^{-6}$, $0.5 \times 10^{-5}$
and $10^{-5}$ from the bottom to the top curve.
As the synchrotron cooling is increased, the inner critical point ceases to exist and one would expect
a shock free solution.
There is an intermediate range of $r$ which cannot be a critical point.
Fig. 2 considers bremsstrahlung cooling case without any synchrotron cooling to show
the effect of varying the specific angular momentum ($\lambda$) of the flow.
The value of $\lambda$ for the flow,
in the units of $2GM/c$,
is increased as 1.6, 1.65, 1.7, 1.8, 1.9 and 2.0 from the top to the bottom curve.

\newpage
\hskip -2cm
\begin{picture}(4,450){
\epsfxsize=14cm
\epsfysize=21cm
\epsfbox{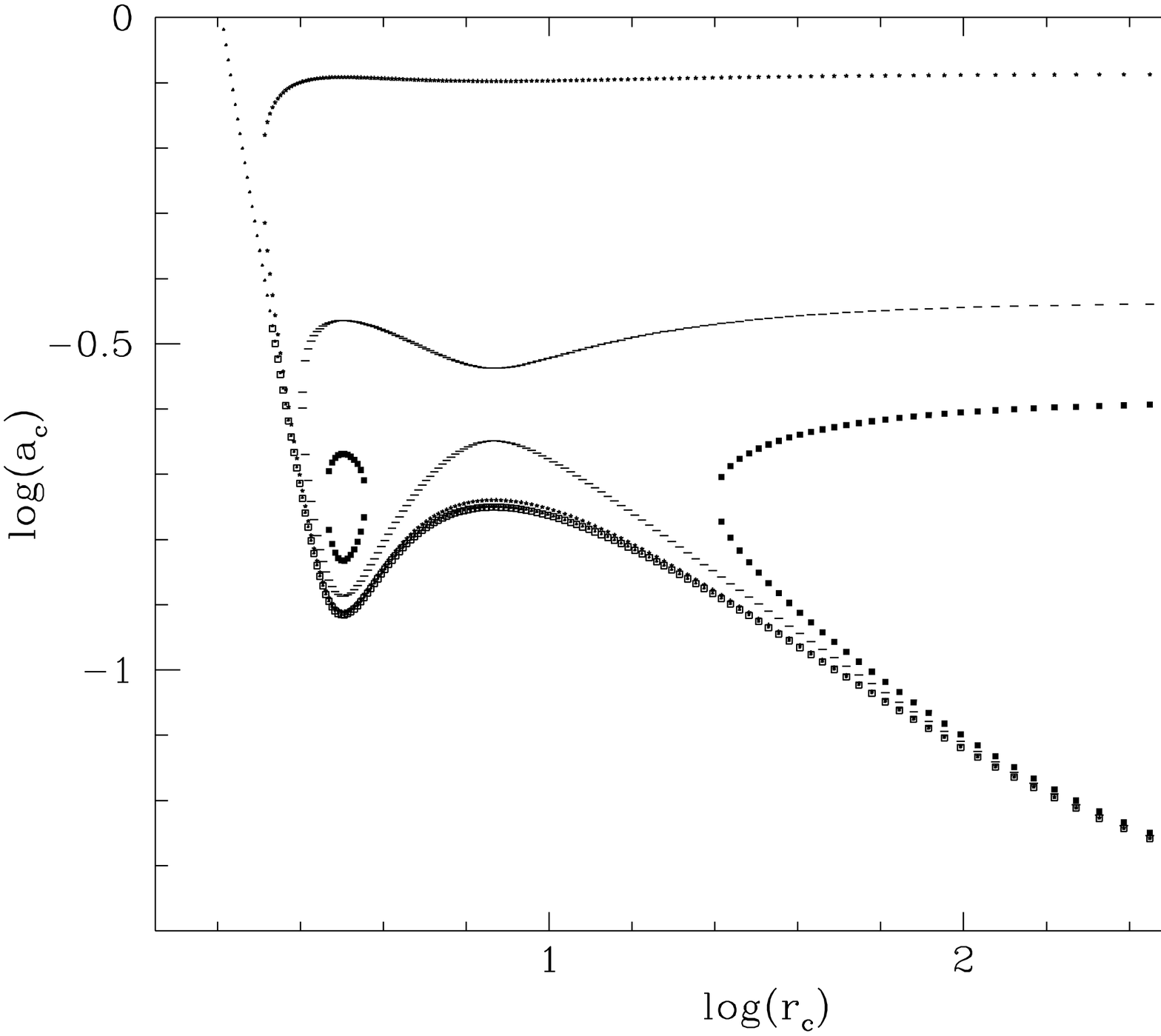} }
\end{picture}
\vspace{1cm}
\\ Fig. 1: Plot of log($a_c$) $vs$ log($r_c$) for accretion rate=1 in Eddington units,
$\lambda$=1.8 in the units of $2GM/c$ and
different values of efficiency of
synchrotron cooling. The efficiency factor of synchrotron process is increased as 0, $10^{-7}$,
$10^{-6}$, $0.5 \times 10^{-5}$ and $10^{-5}$ from bottom to top curve. The efficiency of
bremsstrahlung process is unity for all cases.

\newpage
\hskip -2cm
\begin{picture}(4,450){
\epsfxsize=14cm
\epsfysize=21cm
\epsfbox{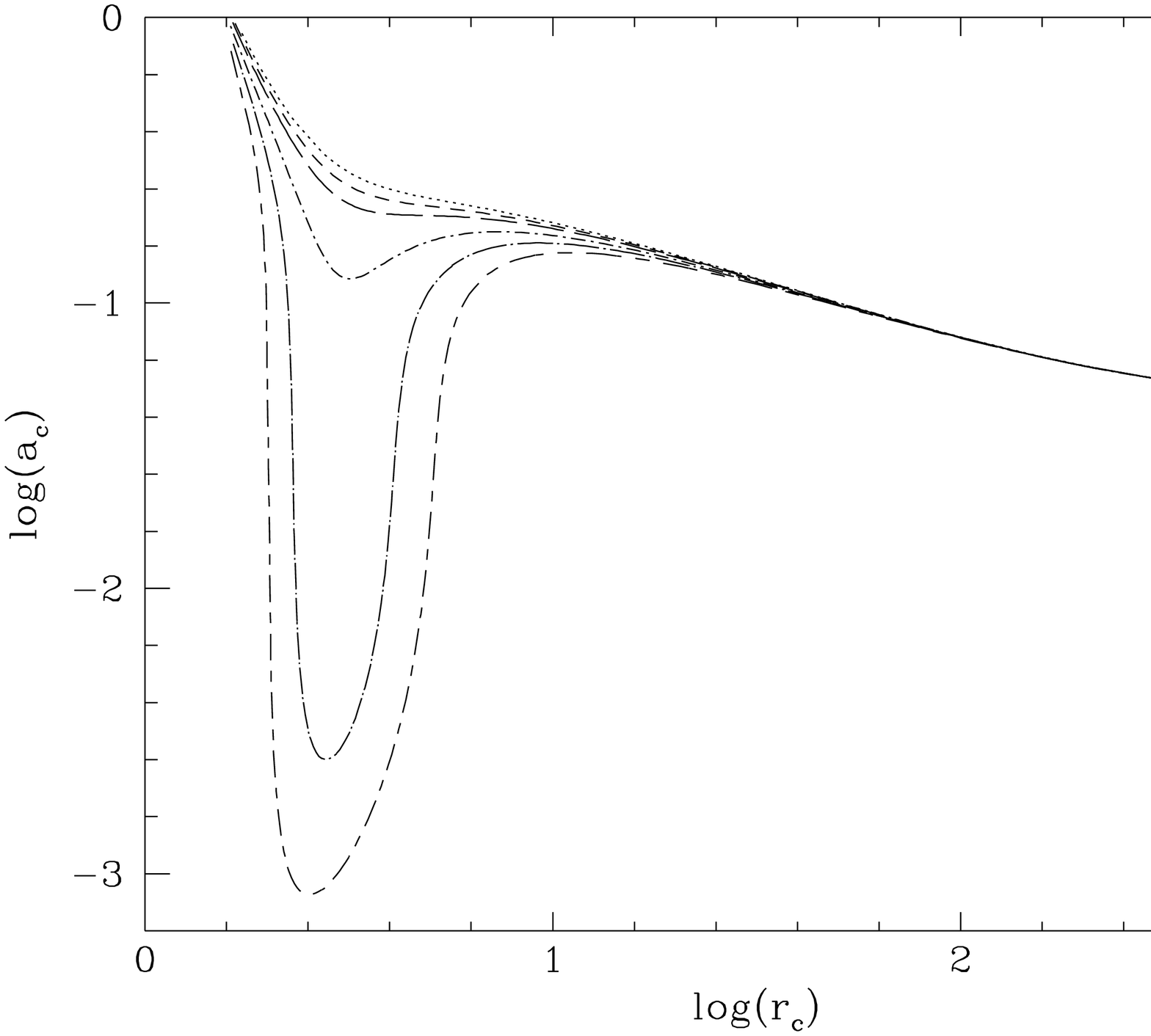} }
\end{picture}
\vspace{1cm}
\\ Fig. 2: Plot of log($a_c$) $vs$ log($r_c$) for accretion rate=1 in Eddington units,
efficiency of bremsstrahlung process is unity and with no synchrotron cooling.
The value of specific angular momentum $\lambda$, in the units of $2GM/c$, is increased as 1.6, 1.65, 1.7, 1.8,
1.9 and 2.0 from top to bottom curve.

\newpage
\noindent {\bf 3.2 Flow topology}

\noindent We choose the accretion rate and specific angular momentum of the flow 
to be the free parameters.
The specific angular momentum is varied in the range (1.6, 2.5), in the units of $2GM/c$ .
It has been suggested that flows with specific angular momentum close to that of marginally stable
and marginally bound orbit
would form steady accretion discs, when the self-gravity of the disc is neglected (Chakrabarti 1996b).
The location of the critical point is chosen to be between $1.5 r_g$ and $1000 r_g$.
In the case where only bremsstrahlung cooling process is considered, the critical point
can be any one of the five types as shown in Fig. 3. They are named as X or Alp or
plA or lA or x type. X looks like the English alphabet X, Alp like Greek alphabet alpha which opens
towards infinity, plA is reflected
Alp which opens towards the inner boundary, lA is plA which doesn't reach the inner boundary when $\lambda$ is
high
and x is X which doesn't reach the inner boundary when the accretion rate is high.

Figs. 4 and 5 show the topology of the flow for bremsstrahlung cooling case and Fig. 6 for
the case when both
bremsstrahlung and synchrotron cooling are present. The synchrotron cooling factor is chosen as $10^{-6}$.
Supersonic branches are obtained for the outer critical points and subsonic branches for the inner
critical points. The presence of synchrotron cooling is felt more in the subsonic branches.
There is an intermediate range of $r$ which corresponds to O-type critical point of Chakrabarti (1989).
When both bremsstrahlung and synchrotron cooling are considered the accreting matter at large radial
distance is very cold and the sound speed becomes small.
Hence X, Alp and x type critical point ceases to exist.
If the assumption that the efficiency of synchrotron cooling is same at all
radial distances is relaxed, then the flow could start as a subsonic flow far away.
The solution remains exactly the same if the mass of central black
hole is changed from $10^8M_{\odot}$ to
$14M_{\odot}$, if the accretion rate and $\lambda$ are the same in the scaled units.
But there will be a difference due to the value of
$\gamma$ changing from 5/3 to 4/3.

\vspace{-.435cm}

\hskip -3cm
\begin{picture}(400,645){
\epsfxsize=19.5cm
\epsfysize=28cm
\epsfbox{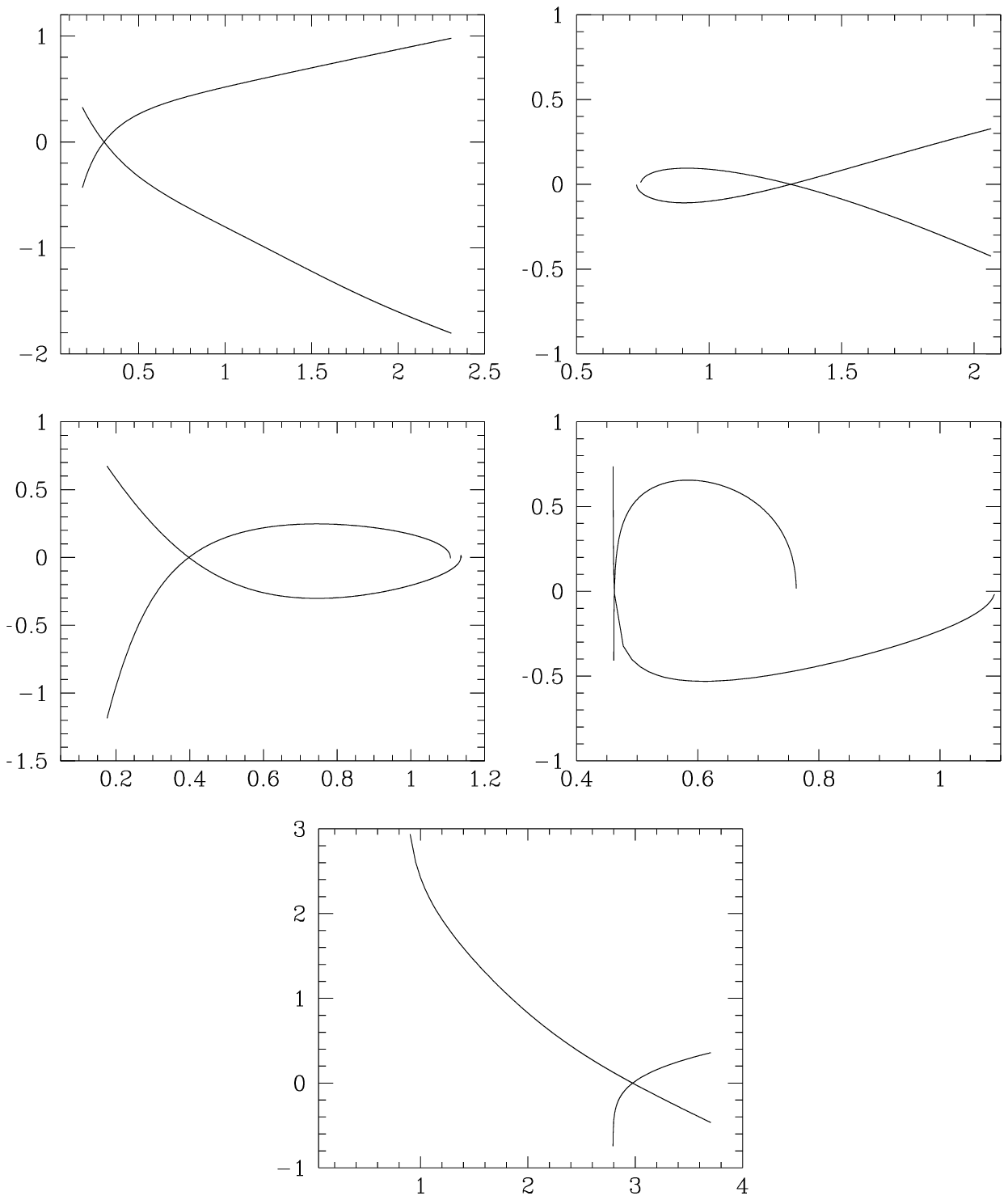} }
\end{picture}
\vspace{-5cm}
\\ Fig. 3: Plot of log(Mach number) $vs$ log(Radial distance) of different components,
X, Alp, plA, lA and x,
of
the topology, for bremsstrahlung cooling case.

\newpage
\hskip -2cm
\begin{picture}(4,585){
\epsfxsize=12cm
\epsfysize=18cm
\epsfbox{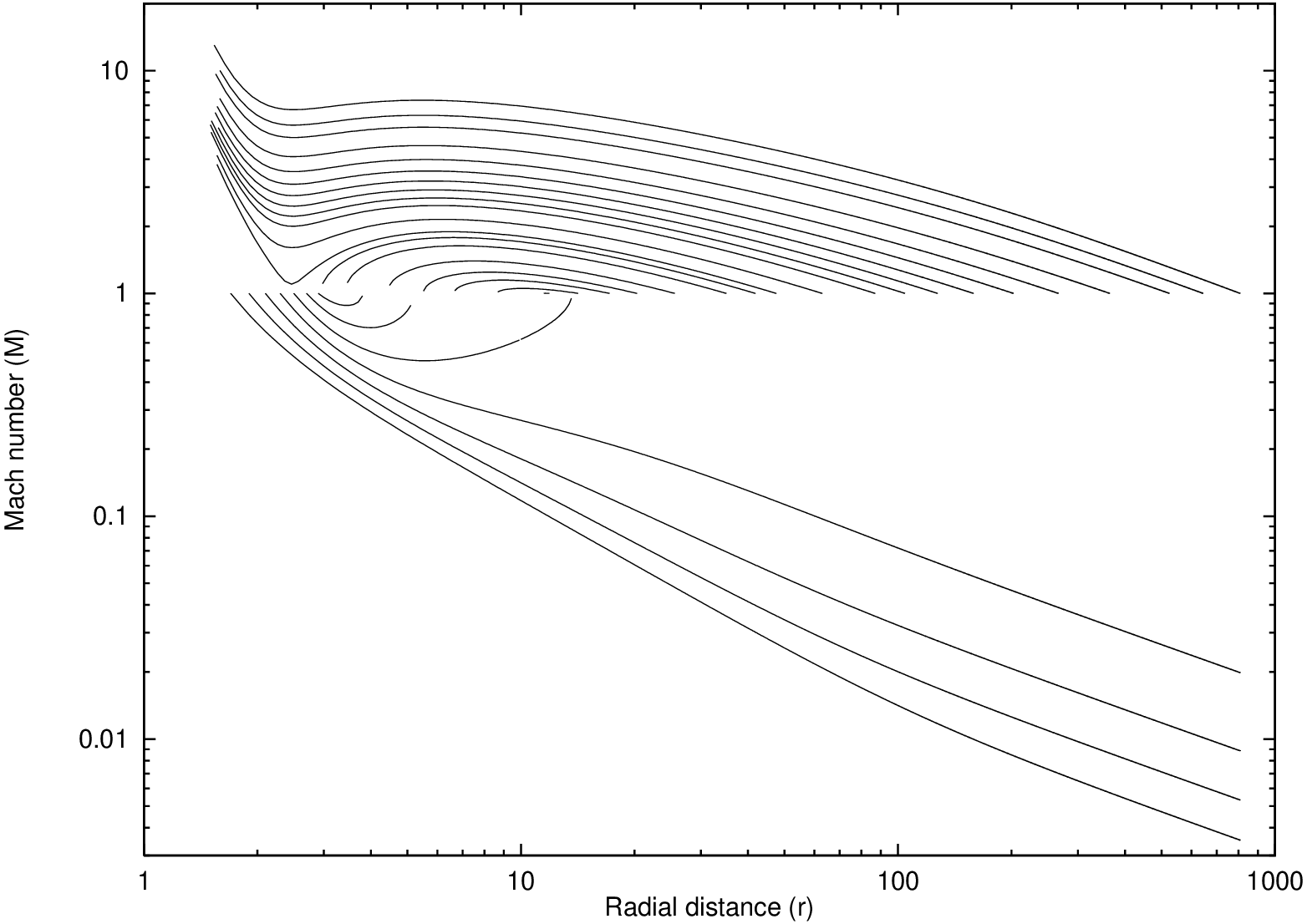}
}
\end{picture}
\vspace{-3cm}
\\ Fig. 4: Mach number $vs$ radial distance of X plA - Alp X topology for accretion rate=1.0
and $\lambda$=1.8. The outermost (supersonic) branches are of X type and it changes to Alp type
as one moves in. There is intermediate range of $r$ which cannot be a critical point and
it is denoted by a -. Further moving in the critical point changes to plA and finally to X
type again. Hence this topology is named as X plA - Alp X.

\newpage
\hskip -2cm
\begin{picture}(4,585){
\epsfxsize=12cm
\epsfysize=18cm
\epsfbox{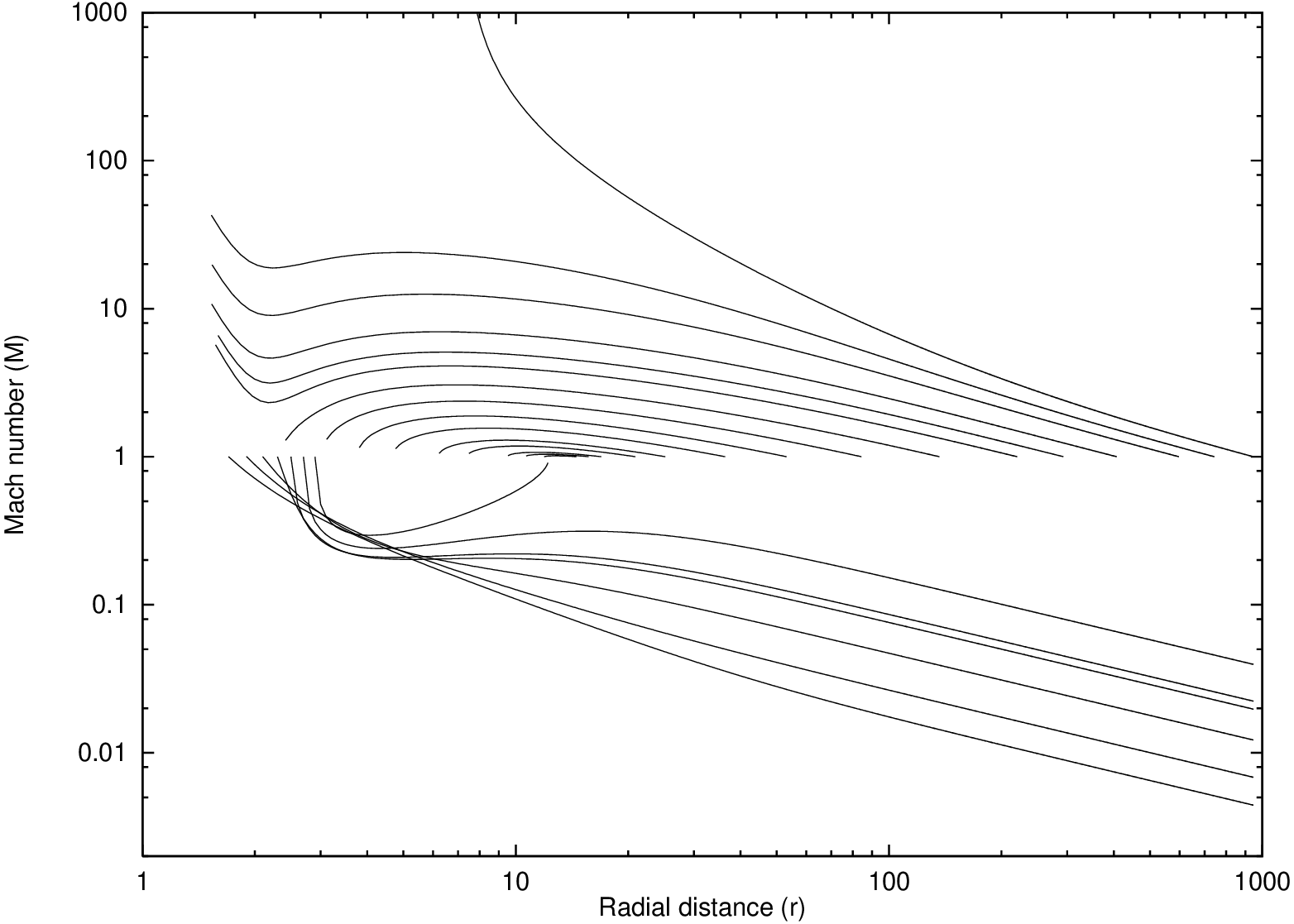}
}
\end{picture}
\vspace{-3cm}
\\ Fig. 5: Mach number $vs$ radial distance of X lA - Alp X x topology for accretion rate=5.0
and $\lambda$=1.9. Note that the outermost (supersonic) branch is of x type, that the flow is
cold enough for the value of sound speed to drop and hence become highly supersonic as it
approaches the event horizon.

\newpage
\hskip -2cm
\begin{picture}(4,585){
\epsfxsize=12cm
\epsfysize=18cm
\epsfbox{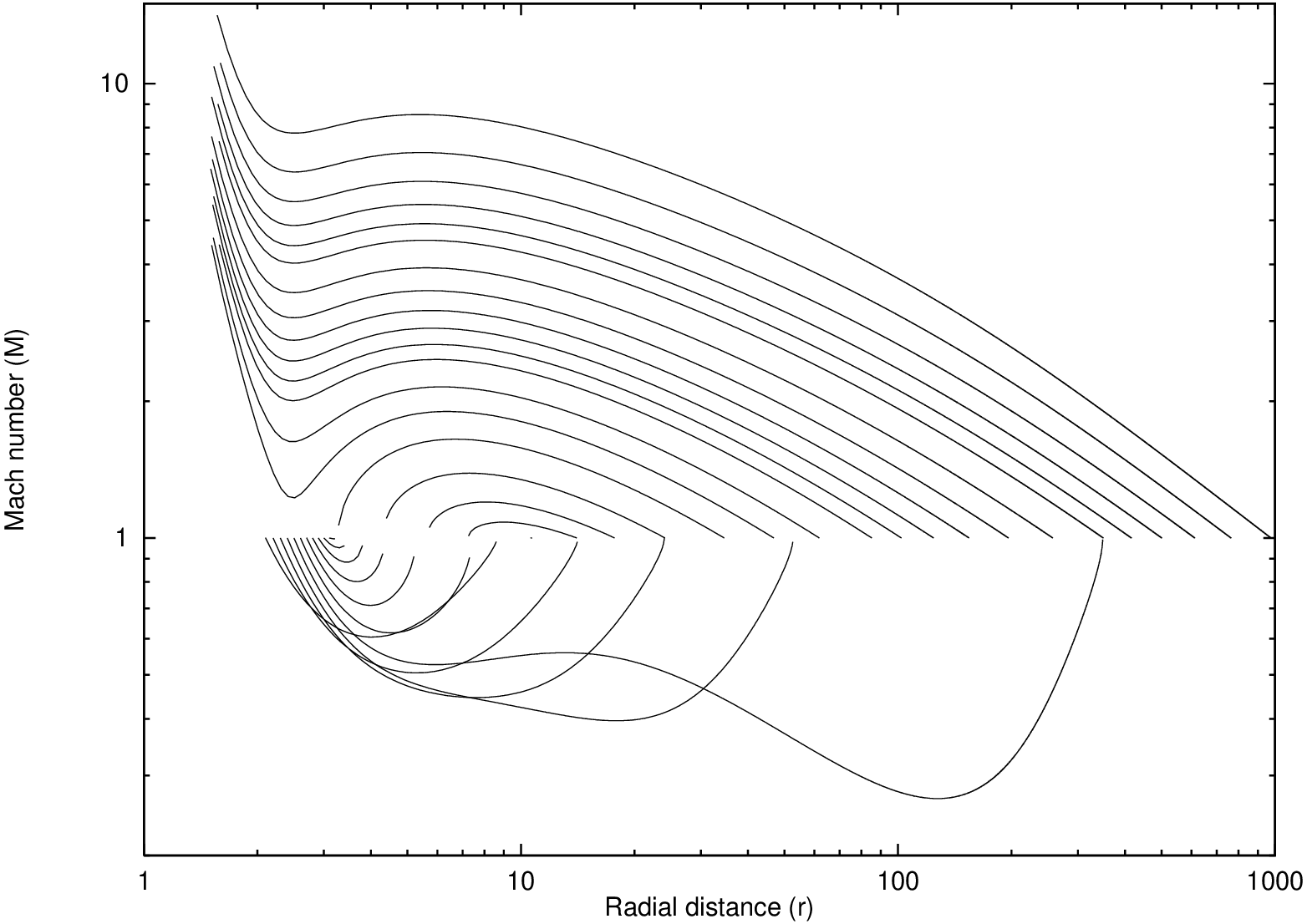}
}
\end{picture}
\vspace{-3cm}
\\ Fig. 6: Mach number $vs$ radial distance topology for accretion rate=1.0 and $\lambda$=1.8 in presence of
synchrotron cooling. The efficiency factor of synchrotron cooling is $10^{-6}$ and full
bremsstrahlung cooling is used. The effect of
synchrotron cooling is felt more in subsonic branches. This is same as Fig. 4 except for
the presence of synchrotron cooling.

\newpage
\noindent {\bf 4. PARAMETER DEPENDENCE OF SHOCK FORMATION IN THE FLOW}

\noindent Mach number is defined as the ratio of the flow velocity to the velocity 
of sound. If the Mach number is
less than unity the flow is called subsonic and if greater than unity it is supersonic.
Flow is called transonic if it makes a transition from subsonic to supersonic flow or vice versa
either continuously or discontinuously.
The point of continuous transition is called the sonic point and the location of discontinuous
transition is called shock.

 The fluid has to be compressible for it to be supersonic. Hence a shock is not possible in an
incompressible fluid. Any perturbation that occurs in a real flow, moves with the medium
and relative
to the medium at the speed of sound. As the perturbation propagates, it steepens due to non-linearity and produces
a shock (shock is  basically a sharp discontinuity).
It is possible for the flow variables to change discontinuously without violating
the conservation laws of mass, momentum and energy.
The strength of shock is defined as the ratio of preshock and postshock quantities. For a weak shock
the discontinuity is small. For moderately strong shock, the discontinuity occurs in
a very thin layer.

Rankine-Hugoniot shock is adiabatic and there is no dissipation of energy at the shock but there is
an increase in the entropy. The velocity jumps from supersonic to subsonic value and the postshock
temperature is higher. Sound speed and thickness of the flow increases.
In an isothermal shock, the mass and momentum are conserved. The energy
dissipated at the shock is radiated away.
The temperature of the preshock and
postshock matter is the same as the cooling process is assumed to be very efficient.
Sound speed and thickness of flow does not change across
the shock.
For isentropic shock, mass and momentum are conserved and specific entropy of the preshock and
postshock matter is the same. There is loss of specific energy which is radiated at the shock location.

The foremost criterion while handling fluid dynamical problems is stability. Whether the flow
will form a shock or not should be decided by stability. Supersonic flows are susceptible to
shocks.
The conditions for an infinitesimally thin and non-dissipative (Rankine-Hugoniot) shock are,

\noindent (a) Pressure balance condition,
$$
p_1 + \rho _1 {v_1}^2 = p_2 + \rho _2 {v_2}^2,
\eqno(9a)
$$

\noindent and,

\noindent(b) Energy flux condition, which reduces to,
$$
{1 \over 2} {v_1}^2 + {{a_1}^2 \over \gamma -1}
= {1 \over 2}  {v_2}^2 + {{a_2}^2 \over \gamma -1},
\eqno(9b)
$$
where, the suffixes 1 and 2 refer to pre-shock and post-shock quantities of the flow respectively.

The Rankine-Hugoniot shock conditions are not satisfied for all values of accretion rate and
specific angular momentum. The region of the parameter space that will form accretion or wind,
with or without shocks, in the case of energy dissipationless flow, has been studied in
Chakrabarti (1989). The parameter space of accretion rate and specific angular
momentum is divided into four main regions, namely no shock in accretion (NSA),
shock in accretion (SA), shock in wind (SW) and no shock in wind (NSW).

In the case of accretion, the entropy corresponding to the inner sonic point is higher
than that of the outer sonic point and the opposite is true for a wind solution.
In the non-dissipative flow, as in Chakrabarti (1989), for a chosen value of accretion rate and
specific angular momentum,
there are unique values for inner and outer critical points.
When we allow the flow to have some dissipation through bremsstrahlung and synchrotron radiation processes, for a chosen accretion
rate and specific angular momentum, there is a range
for outer and inner critical points, whose specific energies are different. But
the specific energy at inner critical point is less than that of the outer, when it forms a shock.

When Rankine-Hugoniot shock conditions, equations 9(a-b), are satisfied the flow could
make a transition from supersonic to subsonic branch and enter the horizon supersonically.
In the case where there is no energy dissipation, shock forms at two
locations inbetween the inner and the outer critical points. Due to energy dissipation both the
shocks may approach each other or one of the shocks
may disappear. Also there are more than one outer-inner branch combination that could form a valid shock solution.
Inner $vs$ outer critical points for which the shocks form are shown in Fig. 7(a-b) and Fig.
7(c-d).
The Figs. 8 and 9 show the shock location $vs$ outer and inner critical point location respectively, 
for valid shock solutions.
In some of the cases, especially for $\lambda$=1.8,
both the inner shock and outer shock are distinctly visible.
 Due to presence of synchrotron cooling, the outer shock location is seen to move in. The same
effect is seen in the context of white dwarf (see Chanmugam, Langer and Shaviv 1985, Table 1).
There is a range of $r_{cin}$ and $r_{cout}$ which can form a shock.
Among the different possible outer and inner branch combinations that form a
shock solution, the actual flow may choose one of them decided by the stability and the boundary
conditions.

 The code which is used for finding the parameter dependence of shock formation in accretion flow is based
on the following `flow chart'.

 (i) the parameters for the outer branch (accrate, lambda, $a_{cout}, r_{cout},$ zefac, 
\indent synfac) are read
from an input file

 (ii) obtain the range of $r$ that cannot be a critical point and decide the range for 
\indent $r_{cin}$ and $r_{cout}$

 (iii) supersonic branch of outer critical point is obtained

\newpage
\begin{picture}(4,250){
\epsfxsize=10.5cm
\epsfysize=12cm
\epsfbox{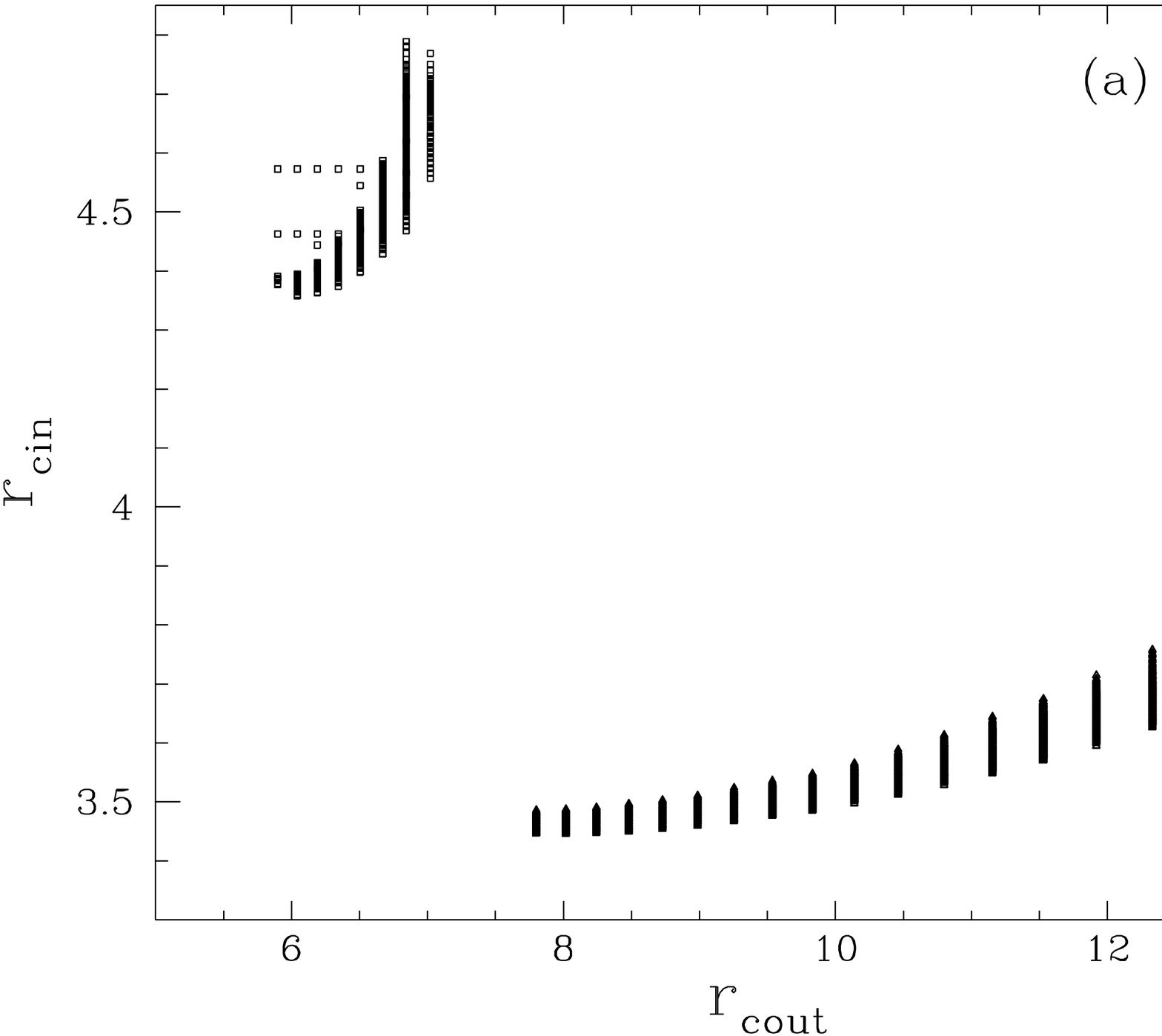} }
\end{picture}

\begin{picture}(4,250){
\epsfxsize=10.5cm
\epsfysize=12cm
\epsfbox{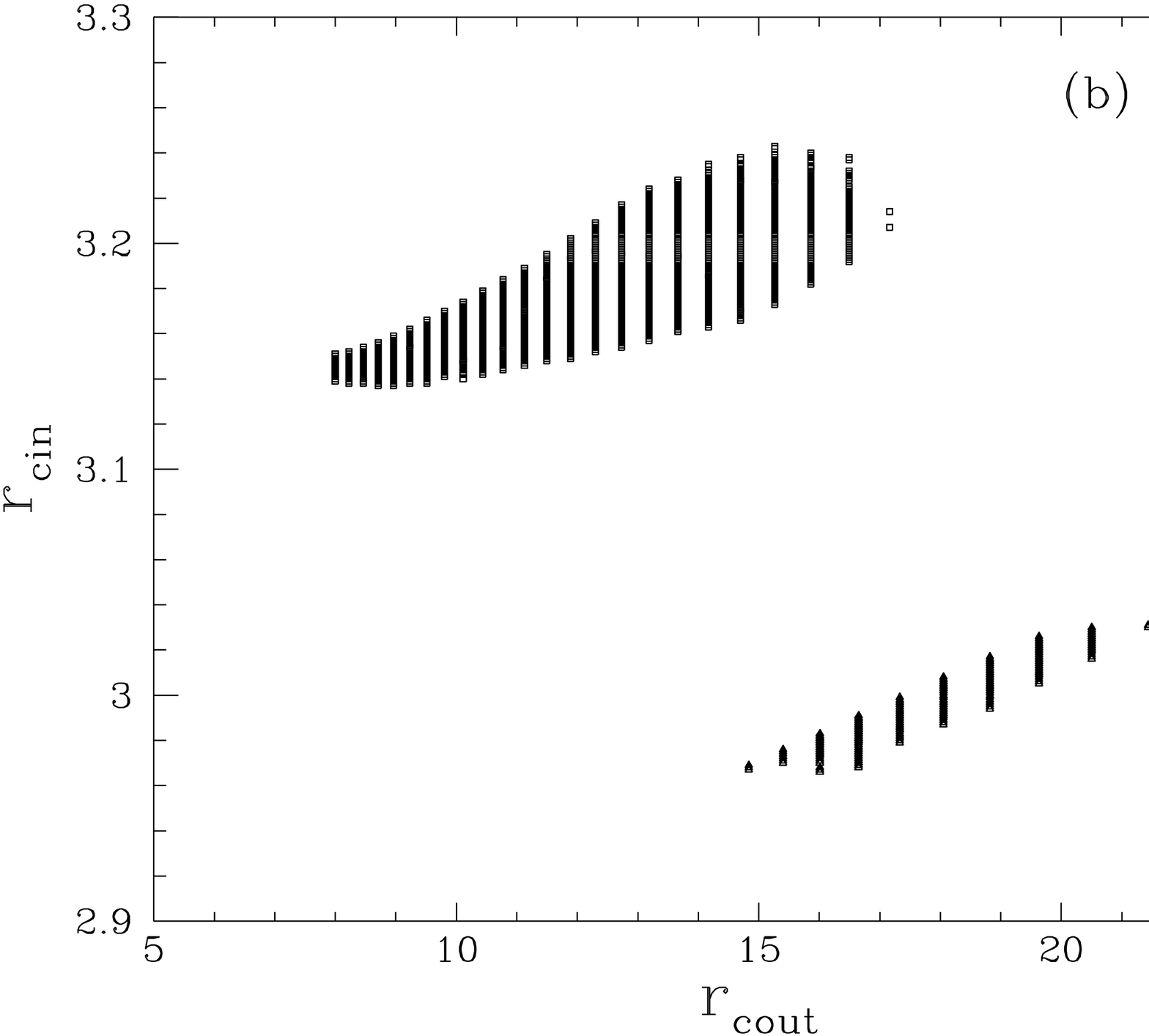} }
\end{picture}
\vspace{.8cm}
\\ Fig. 7(a-b): An example for $r_{cin}$ $vs$ $r_{cout}$ that can form a shock for (a) $\lambda$=1.65
and
(b) $\lambda$=1.7. The upper set of points in both (a) and (b) are for the case of synchrotron
cooling factor $10^{-6}$
and lower ones are for synchrotron cooling factor $0$. The bremsstrahlung cooling factor
is unity and the accretion rate=1.0 in Eddington units.

\newpage
\begin{picture}(4,250){
\epsfxsize=10.5cm
\epsfysize=12cm
\epsfbox{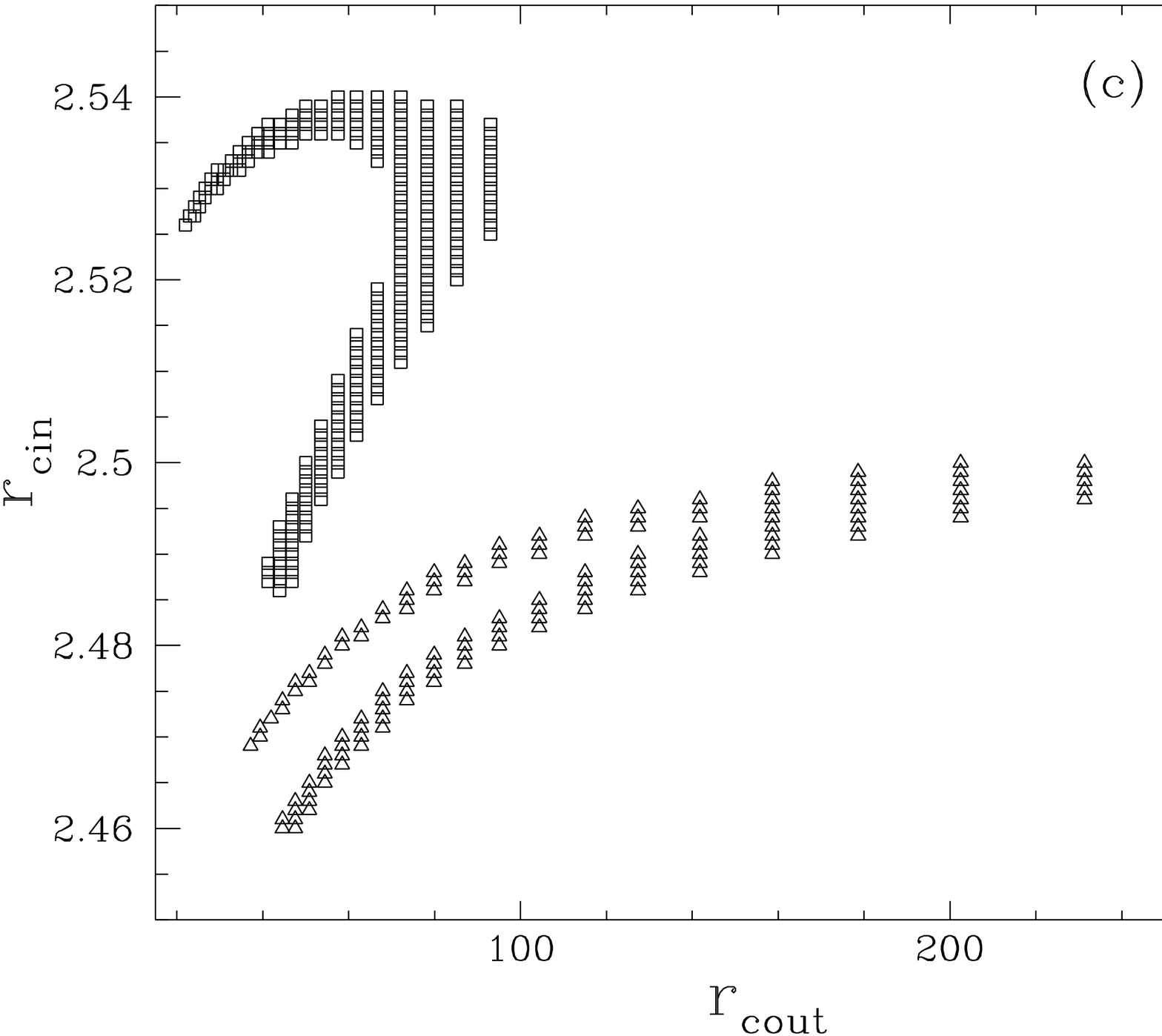} }
\end{picture}

\begin{picture}(4,250){
\epsfxsize=10.5cm
\epsfysize=12cm
\epsfbox{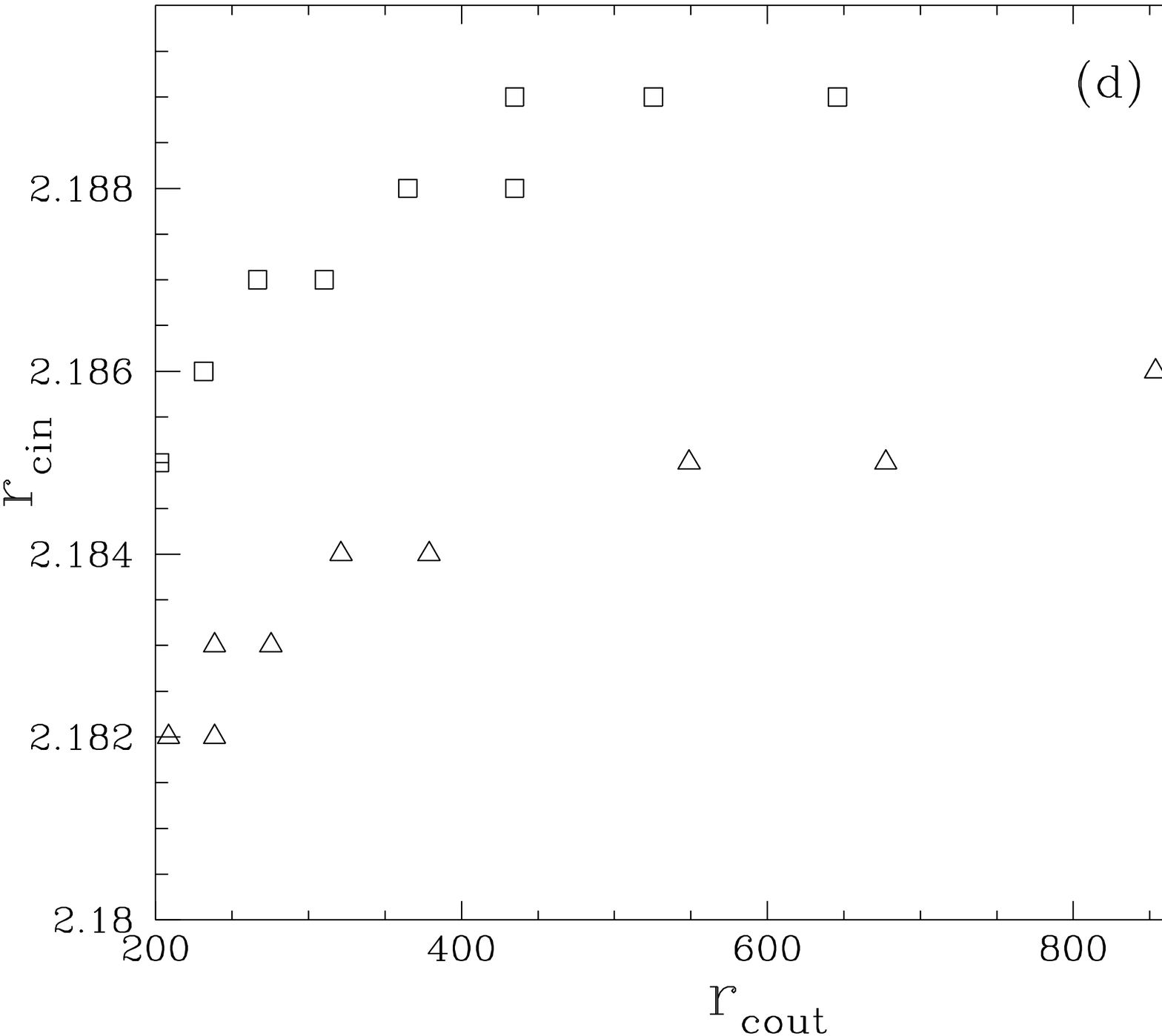} }
\end{picture}
\vspace{.8cm}
\\ Fig. 7(c-d): An example for $r_{cin}$ $vs$ $r_{cout}$ that can form a shock for (c) $\lambda$=1.8 and
(d) $\lambda$=1.9. The upper set of points in both (c) and (d) are for the case of synchrotron
cooling factor $10^{-6}$
and lower ones are for synchrotron cooling factor $0$. The bremsstrahlung cooling factor
is unity and the accretion rate=1.0 in Eddington units.

\newpage
\begin{picture}(4,250){
\epsfxsize=10.5cm
\epsfysize=12cm
\epsfbox{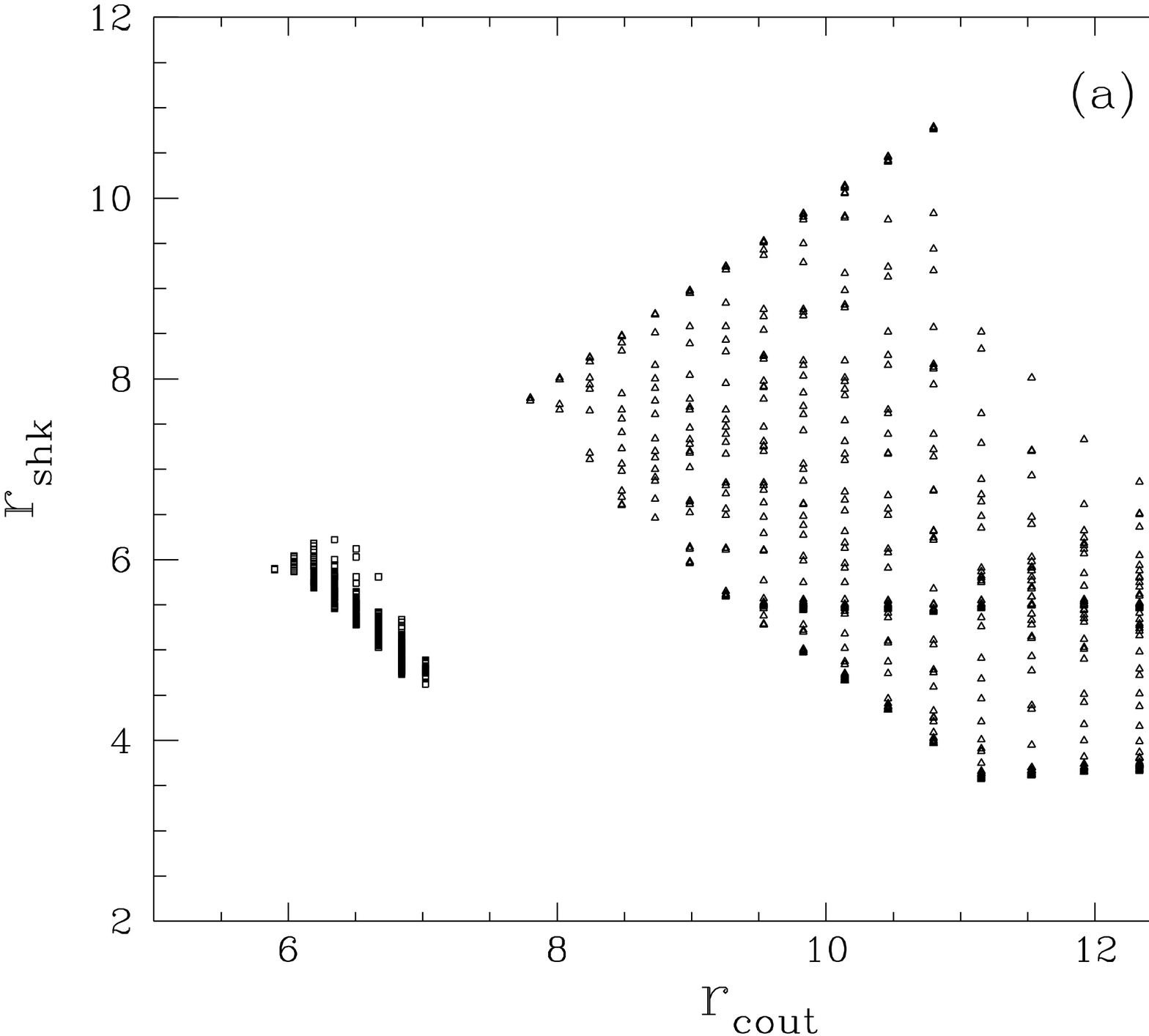} }
\end{picture}

\begin{picture}(4,250){
\epsfxsize=10.5cm
\epsfysize=12cm
\epsfbox{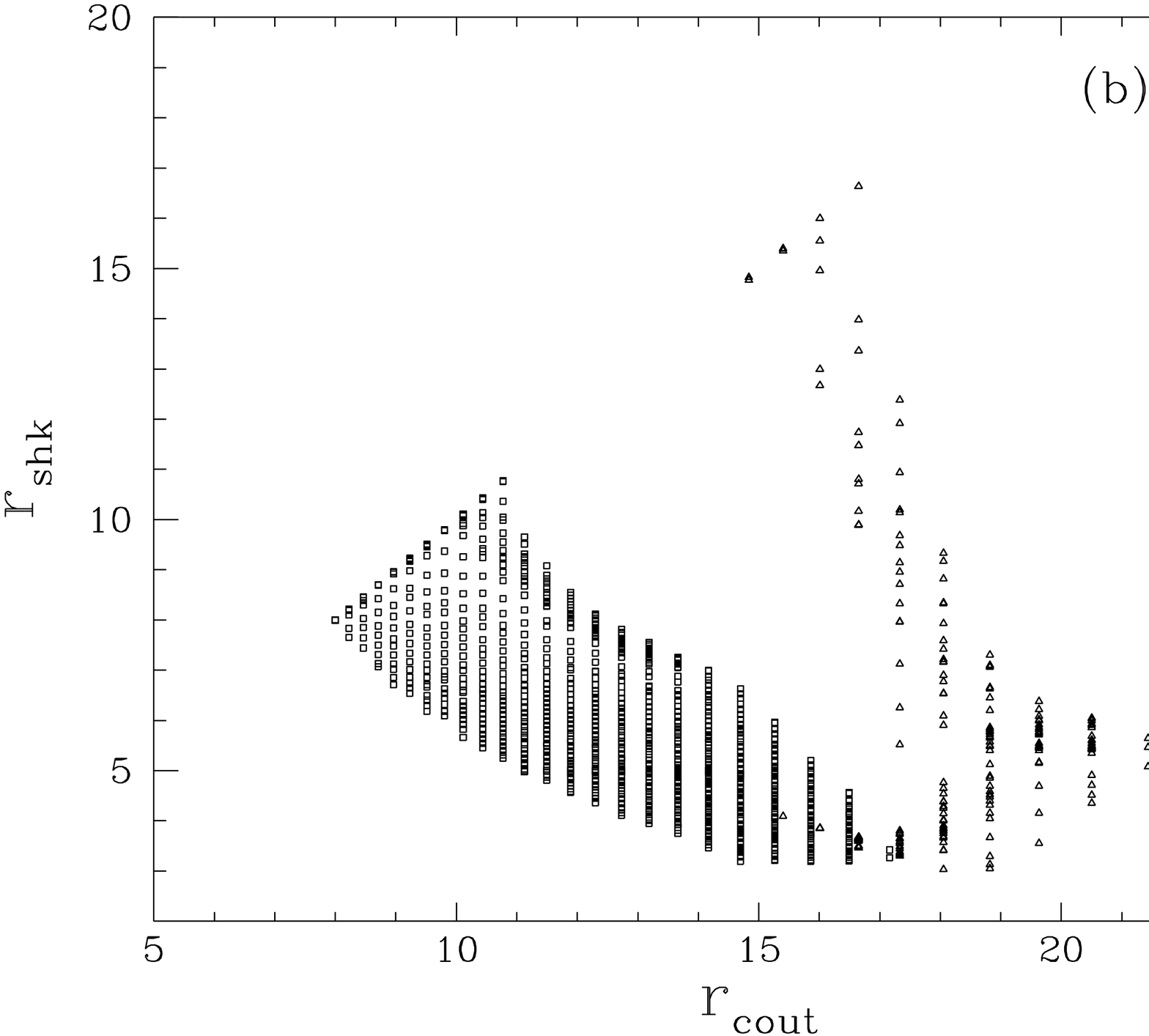} }
\end{picture}
\vspace{1.2cm}
\\ Fig. 8(a-b): The parameters are same as that of Fig. 7(a-b). The possible shock locations
($r_{shk}$) for a chosen outer critical point location ($r_{cout}$) is shown. Each
shock location corresponds to one inner critical point location ($r_{cin}$).
The left set of points are for the case of synchrotron cooling factor $10^{-6}$.

\newpage
\begin{picture}(4,250){
\epsfxsize=10.5cm
\epsfysize=12cm
\epsfbox{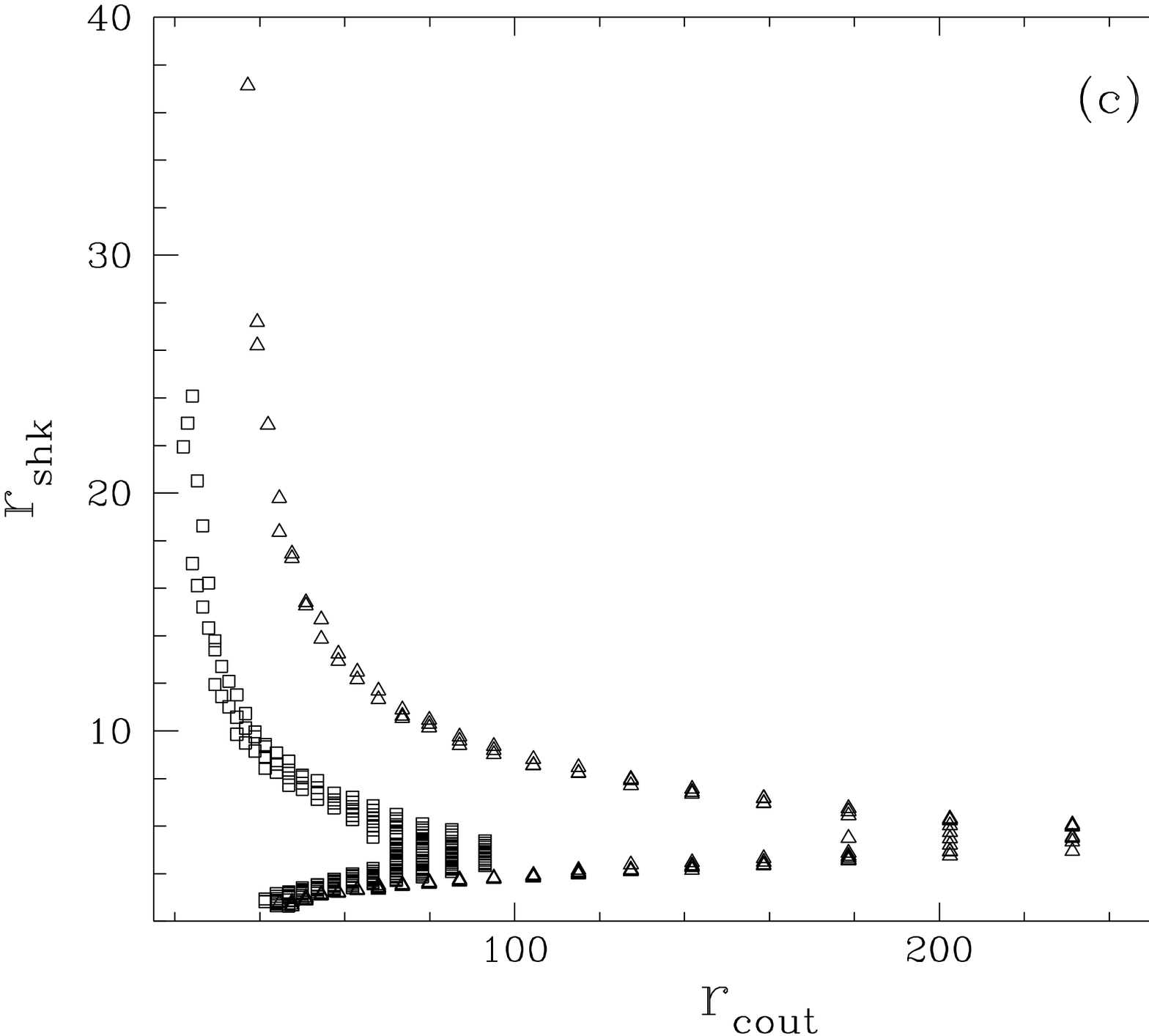} }
\end{picture}

\begin{picture}(4,250){
\epsfxsize=10.5cm
\epsfysize=12cm
\epsfbox{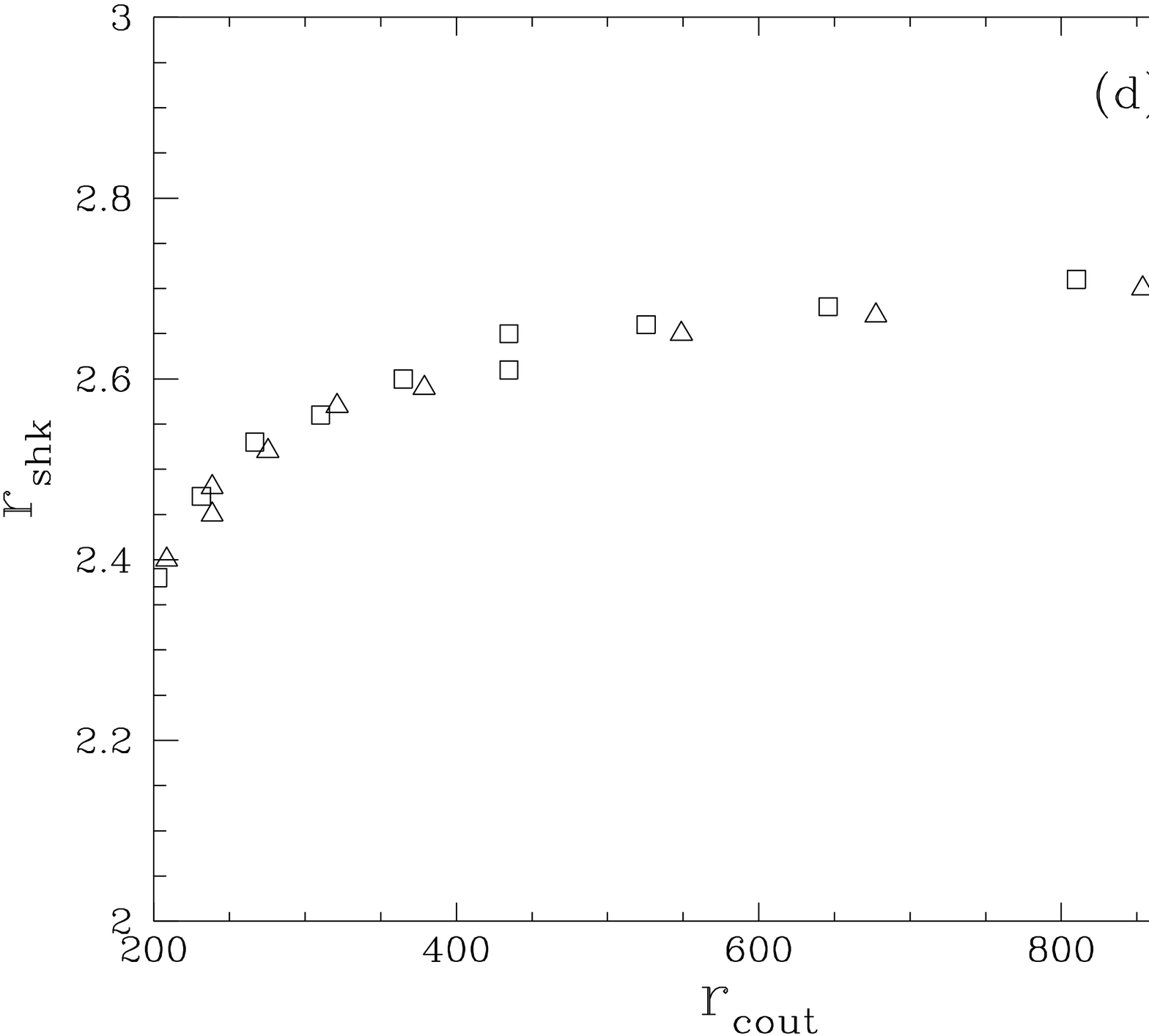} }
\end{picture}
\vspace{.8cm}
\\ Fig. 8(c-d): The parameters are same as that of Fig. 7(c-d). The possible shock locations
($r_{shk}$) for chosen $r_{cout}$ is shown.
For $\lambda=1.8$ case both inner and outer shock locations are distinctly visible. Both
inner and outer shock merge together as $r_{cout}$ moves out. Only inner shock is possible for
the case of $\lambda=1.9$. Left set of points are for the case of synchrotron cooling factor $10^{-6}$.

\newpage
\begin{picture}(4,250){
\epsfxsize=10.5cm
\epsfysize=12cm
\epsfbox{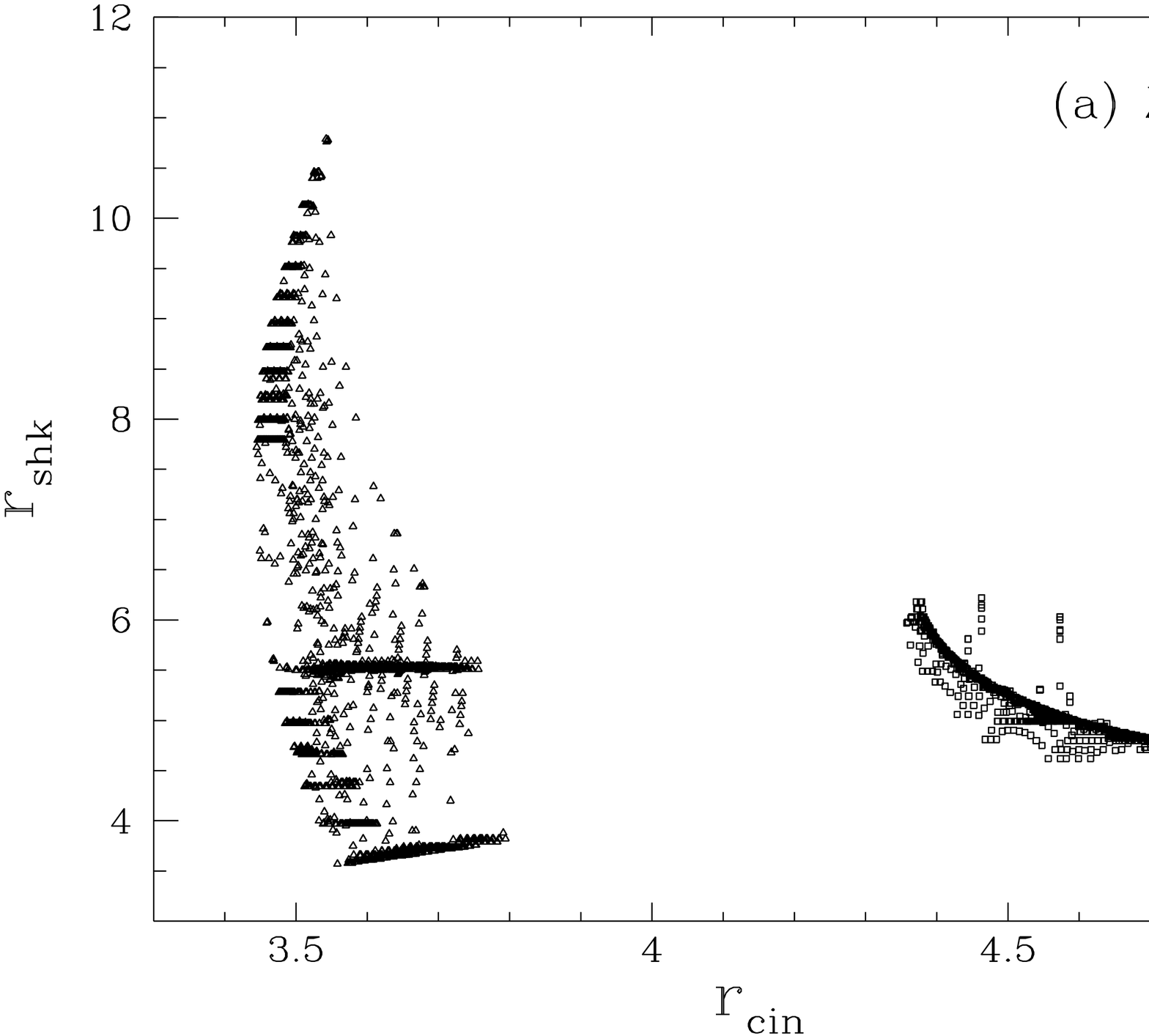} }
\end{picture}

\begin{picture}(4,250){
\epsfxsize=10.5cm
\epsfysize=12cm
\epsfbox{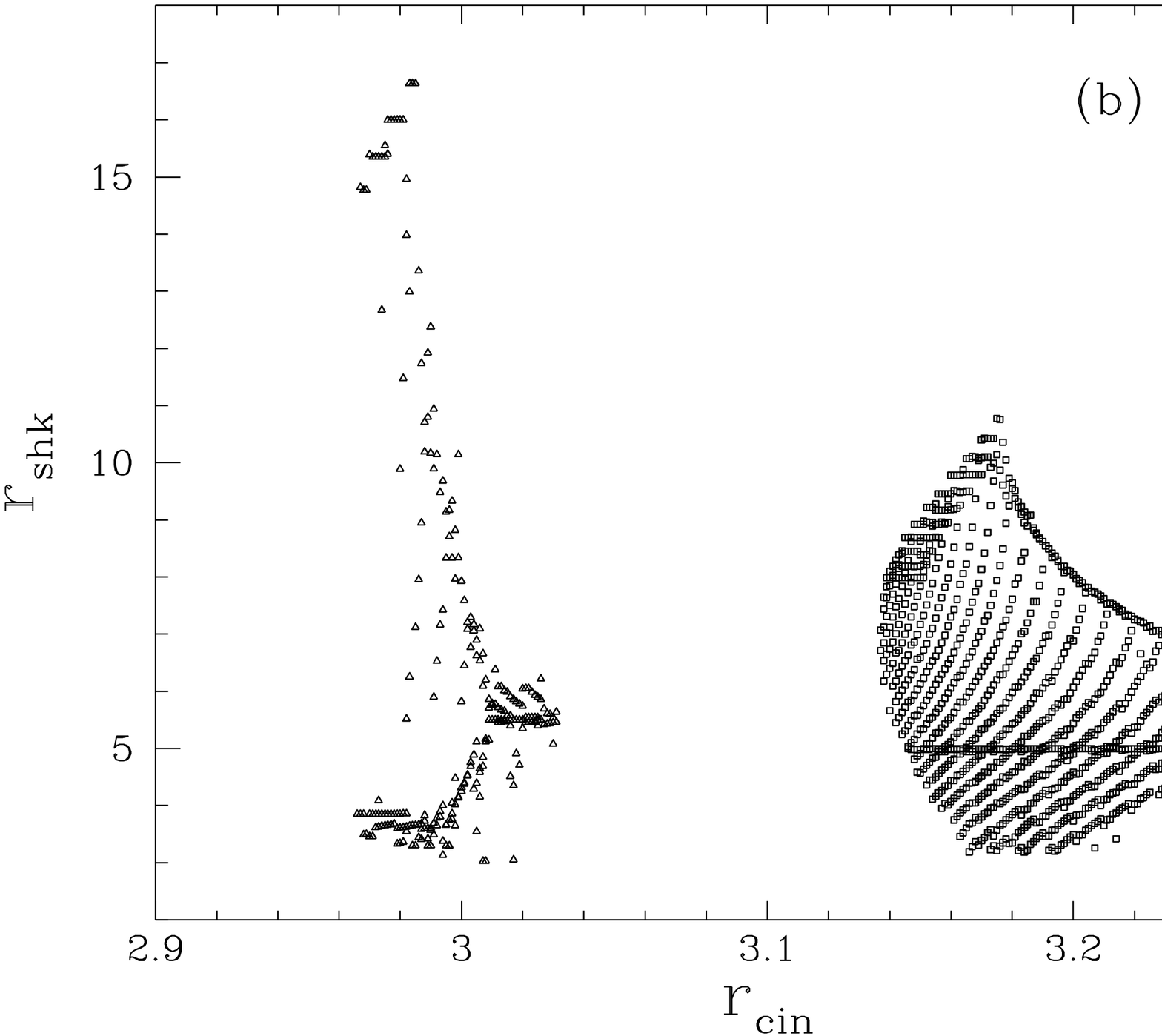} }
\end{picture}
\vspace{1.2cm}
\\ Fig. 9(a-b): The parameters are same as that of Fig. 7(a-b). For a chosen inner critical
point location ($r_{cin}$), different shock locations ($r_{shk}$) are possible. The right
set of points correspond to the case of synchrotron cooling factor $10^{-6}$.

\newpage
\begin{picture}(4,250){
\epsfxsize=10.5cm
\epsfysize=12cm
\epsfbox{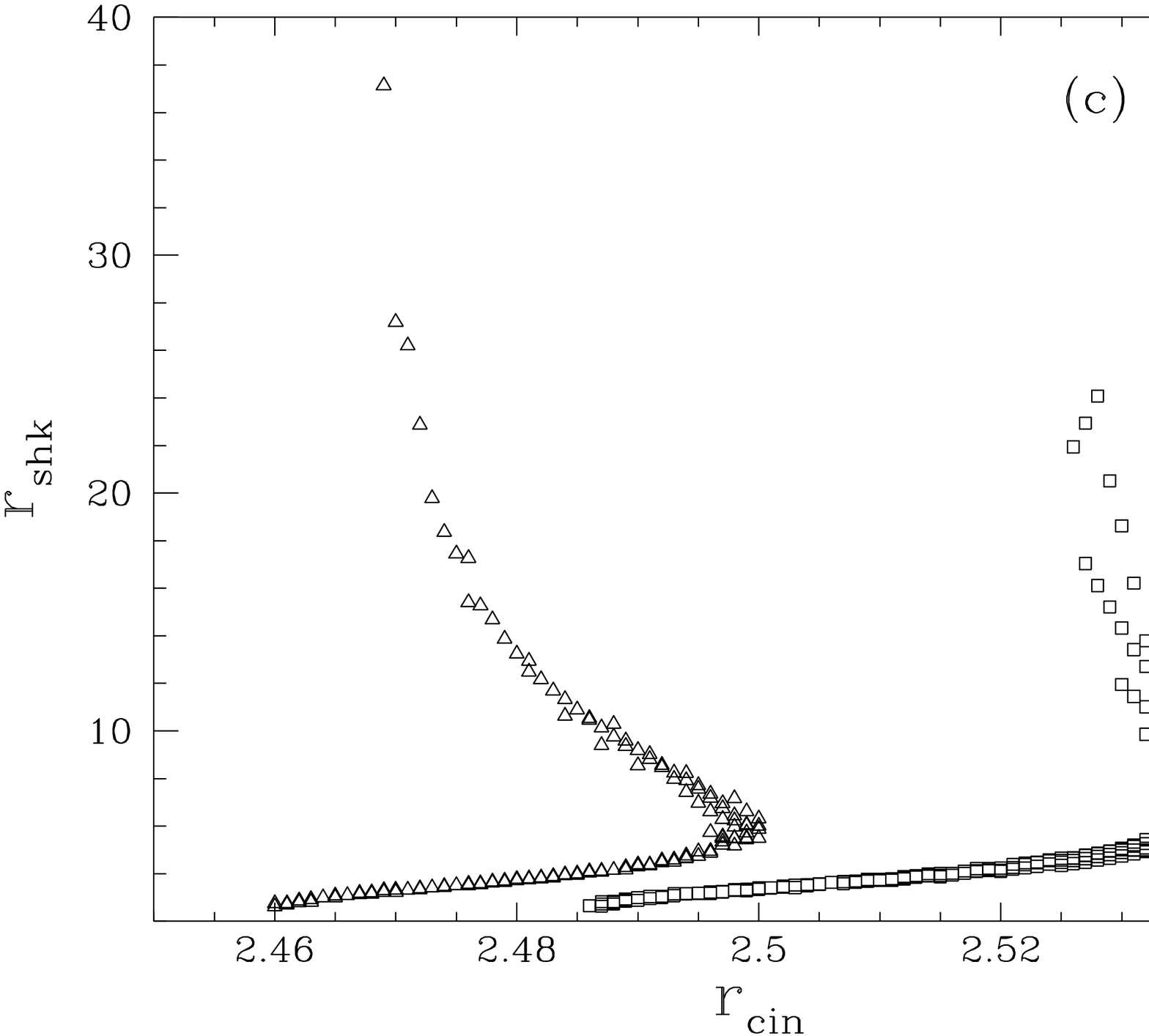} }
\end{picture}

\begin{picture}(4,250){
\epsfxsize=10.5cm
\epsfysize=12cm
\epsfbox{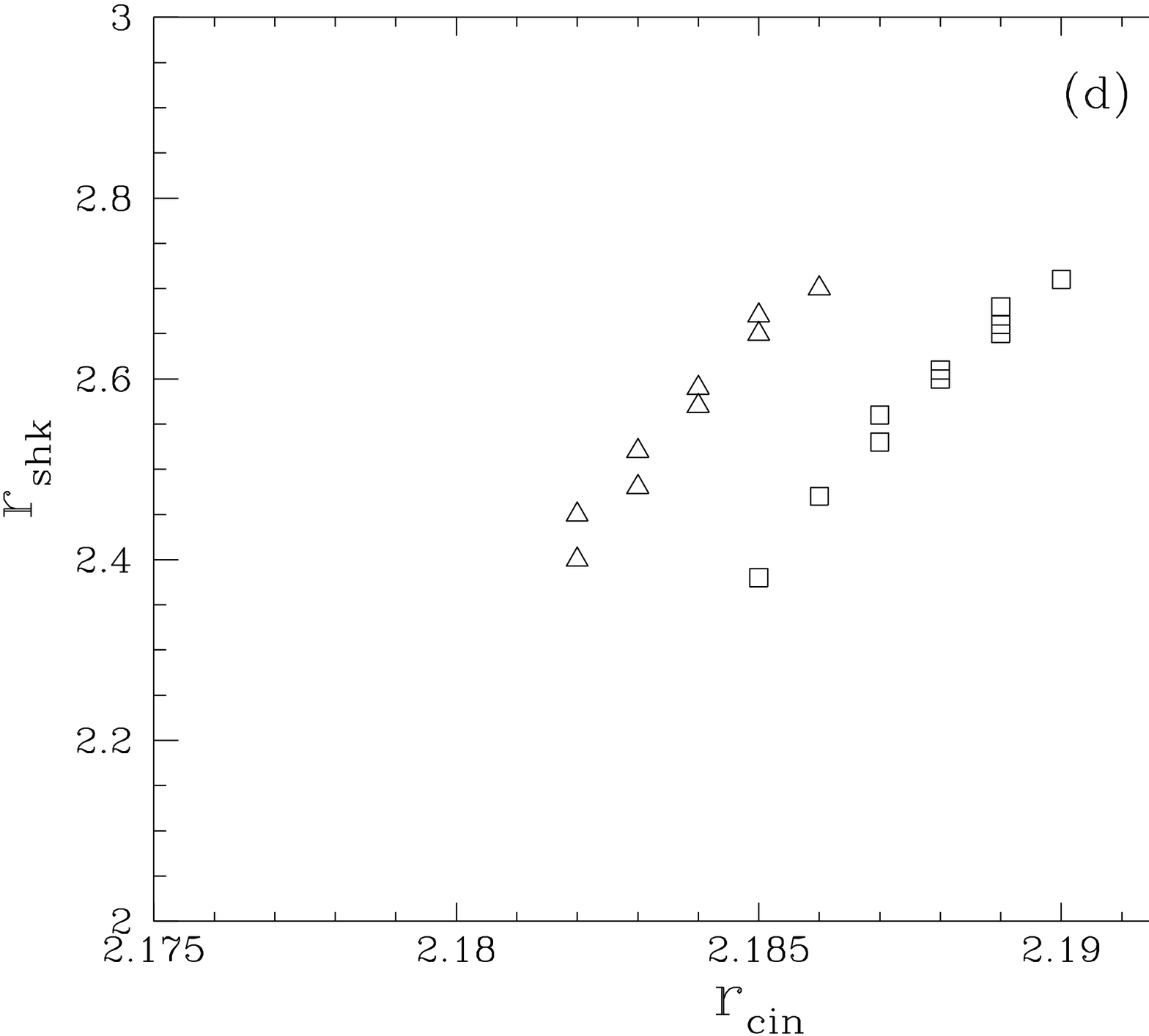} }
\end{picture}
\vspace{1.2cm}
\\ Fig. 9(c-d): The parameters are same as that of Fig. 7(c-d). Both inner and outer shock
locations are distinctly visible for the case of $\lambda=1.8$. For $\lambda=1.9$ only
inner shock forms. The right
set of points correspond to the case of synchrotron cooling factor $10^{-6}$.

\newpage

 (iv) subsonic branch of inner critical point is obtained

 (v) Rankine-Hugoniot shock conditions are checked to find a shock connecting
\indent the supersonic branch and
a subsonic branch

 (vi) for a chosen $r_{cout}, r_{cin}$ is incremented in a loop and all the inner branches
\indent that shock
with
a given outer branch are found

 (vii) loop for $r_{cout}$ by reading next set of parameters for outer branch

\ \\
\noindent {\bf 5. DISCUSSION AND CONCLUSIONS}

\noindent When solutions with shock and shock free solutions are possible for a given accretion rate
and $\lambda$, the flow might choose a shock solution as supersonic flows are susceptible to
shocks. The inner region of the disc will be hotter when it forms a shock compared to a shock free
disc. Hence discs with shocks may be a better candidate for explaining the hard X-rays and $\gamma$-rays,
instead of solely relying on advection effects to increase the temperature (Narayan 1997).
The assumption that the efficiency of synchrotron cooling is the same at all radial distances
could be relaxed for more realistic modeling if matter were to start as a subsonic flow far away.
It will be worthwhile to write the dynamical equations in curved space-time
using the general theory of relativity before embarking into additional details of realistic modeling,
as the flow velocities are relativistic
and also the inner boundary conditions can be taken care properly.

\ \\
\noindent{\bf REFERENCES}

\bigskip

\noindent Artemova, I. V., Bjornsson, G., and Novikov, I. D. 1996, ApJ, 461, 565

\noindent Chakrabarti, S. K. 1989, ApJ, 347, 365

\noindent Chakrabarti, S. K. 1990, MNRAS, 243, 610

\noindent Chakrabarti, S. K. 1996a, ApJ, 464, 664

\noindent Chakrabarti, S. K. 1996b, Physics Reports 266, No. 5 \& 6, 229

\noindent Chakrabarti, S. K. and Khanna, R. 1992, MNRAS, 256, 300

\noindent Chakrabarti, S. K. and
 Molteni, D. 1993, ApJ, 417, 671

\noindent Chanmugam, G., Langer, S. H. and Shaviv, G. 1985, 299, L87

\noindent Das, T. K. 2002, ApJ, 577, 880

\noindent Das, T. K. 2003, ApJ, 588, L89

\noindent Das, T. K., and  Sarkar, A. 2001, A\&A, 374, 1150

\noindent Frank, J., King, A. and Raine, D. 1992, Accretion Power in Astrophysics, second 
\indent edition, Cambridge University Press

\noindent Greiner, J., Cuby, J. G., and McCaughrean, M. J. 2001, Nature, 414, 522

\noindent Lang, K. R. 1980, Astrophysical Formulae, second edition, Springer-Verlag

\noindent Longair, M.S. 1994, High Energy Astrophysics Vol. 2, Cambridge University Press.

\noindent Molteni, D.,
 Lanzafame, G. and
 Chakrabarti, S. K. 1994, ApJ, 425, 161

\noindent Molteni, D., Sponholz, H. and Chakrabarti, S. K. 1996, ApJ, 457, 805

\noindent Mukhopadhyay, B. 2002, ApJ, 581, 427

\noindent Mukhopadhyay, B. 2003, MNRAS, 342, 274

\noindent Narayan, R. 1997, in Unsolved problems in Astrophysics, ed. John N. Bahcall and
\indent Jeremiah P. Ostriker, Princeton Univ. Press

\noindent Narayan, R., Kato, S. and Honma, F. 1997, ApJ, 476, 49

\noindent Narayan, R. and Yi, I. 1994, ApJ, 428, L13

\noindent Nowak, M. A. and Wagoner, R. V. 1991, ApJ, 378, 656

\noindent Paczy{\'n}ski, B. and Wiita, P. J. 1980, A\&A, 88, 23

\noindent Press, W. H., Teukolsky, S. A., Vellerling, W. T. and Flannery, B. P. 1992, Numerical
\indent Recipes in Fortran - The Art of Scientific Computing, second edition, Cambridge
\indent University Press

\noindent Shakura, N. I. and Sunyaev, R. A. 1973, A\&A, 24, 337

\noindent Shapiro, S. L.  and Teukolsky, S. A. 1983, Black Holes, White Dwarfs,
and Neutron 
\indent Stars, New York; Wiley

\pagebreak
\eject
\end{document}